\documentclass[useAMS,usenatbib]{mn2e}

\usepackage{graphicx}
\usepackage{natbib}

\title[The role of low-mass star clusters in forming the massive stars in DR 21]{The role of low-mass star clusters in forming the massive stars in DR 21}
\author[V.M. Rivilla, I. Jim\'enez-Serra, J. Mart\'in-Pintado and J. Sanz-Forcada]{V.M. Rivilla$^{1}$\thanks{E-mail:
rivilla@cab.inta-csic.es; ryvendel@gmail.com}, I. Jim\'enez-Serra$^{2}$, J. Mart\'in-Pintado$^{1}$ and J. Sanz-Forcada$^{1}$ 
\\
\\
$^{1}$Centro de Astrobiolog\'ia (CSIC-INTA), Ctra. de Torrej\'on Ajalvir, km. 4, E-28850 Torrej\'on de Ardoz, Madrid, Spain\\
$^{2}$European Southern Observatory, Karl-Schwarzschild-Str. 2, 85748, Garching, Germany}

\begin{document}

\date{Accepted XXXX December XXX. Received XXXX December XX}

\pagerange{\pageref{firstpage}--\pageref{lastpage}} \pubyear{2013}

\maketitle

\label{firstpage}

\begin{abstract}
We have studied the young low-mass pre-main sequence (PMS) stellar population associated with the massive star-forming region DR 21 by using archival X-ray Chandra observations and by complementing them with existing optical and IR surveys. The Chandra observations have revealed for the first time a new highly extincted population of PMS low-mass stars previously missed in observations at other wavelengths. 
The X-ray population exhibits three main stellar density peaks, coincident with the massive star-forming regions, being the DR 21 core the main peak.
The cross-correlated X-ray/IR sample exhibits a radial "Spokes-like" stellar filamentary structure that extends from the DR 21 core towards the northeast.
The near IR data reveal a centrally peaked structure for the extinction, which exhibits its maximum in the DR 21 core and gradually decreases with the distance to the N-S cloud axis and to the cluster center. 
We find evidence of a global mass segregation in the full low-mass stellar cluster, and of an stellar age segregation, with the youngest stars still embedded in the N-S cloud, and more evolved stars more spatially distributed. The results are consistent with the scenario where an elongated overall potential well created by the full low-mass stellar cluster funnels gas through filaments feeding stellar formation. Besides the full gravitational well, smaller-scale local potential wells created by dense stellar sub-clusters of low-mass stars are privileged in the competition for the gas of the common reservoir, allowing the formation of massive stars. We also discuss the possibility that a stellar collision in the very dense stellar cluster revealed by Chandra in the DR 21 core is the origin of the large-scale and highly-energetic outflow arising from this region.

\end{abstract}


\section{Introduction}
\label{intro}

The formation of massive stars is one of the most debated topic in modern Astrophysics. Although they are a key ingredient in the evolution of galaxies, because they inject large amount of energy and turbulence into the interstellar medium, the processes leading to their formation are not fully understood.

Massive stars are usually born in clusters (\citealt{lada03}), suggesting that clusters play an important role in massive star formation. 
Smoothed particle hydrodynamics simulations of massive star forming clumps in a giant molecular cloud carried out by \citet{smith09b} have shown that the formation of massive stars is closely linked to the formation and early evolution of the whole stellar cluster. Recently, \citet{rivilla13a} have pointed out that the presence of dense sub-clusters of low  mass stars in the Orion Nebula Cluster (ONC) and the Orion Molecular Cloud (OMC) may have been key in the formation of massive stars in this region.

Measuring the population and computing the densities of low-mass star clusters in massive star-forming regions is a challenge because they are usually deeply embedded in the parental molecular cloud (visual extinctions of $A_{\rm V}>$15 mag). X-ray observations are particularly useful for studying the obscured population because the high energy photons can deeply penetrate into the cloud despite the high extinction. In addition, the X-rays have the advantage of suffering much less from foreground/background contamination than optical or infrared (IR) studies, thus allowing to carry out a more complete census of the low-mass pre-main sequence (PMS) embedded population in massive star-forming regions.

The DR 21 massive star-forming region is located at 1.50 kpc (\citealt{rygl12}) and belongs to the Cygnus X molecular cloud. It harbors a massive and dense filament-shaped cloud (\citealt{chandler93}, \citealt{davis07}), forming part of a large-scale network of filamentary structures (\citealt{schneider10}). DR 21 exhibits three main regions of massive star formation: i) the DR 21 core, where several ultra-compact (UC) HII regions are detected (\citealt{cyganowski03}); ii) the DR 21(OH) hot molecular core (\citealt{chandler93}); and iii) the FIR 1/2/3 region (\citealt{chandler93,kumar07}). 

A high energetic outflow has been detected in the DR 21 core (\citealt{garden91,smith06,davis07}), whose origin is still unclear. The outflow has been suggested to be powered by a massive protostar with a luminosity $\sim$ 10$^{5-6}$ L$_{\odot}$ (\citealt{garden91}), which corresponds to a zero age main sequence (ZAMS) star with spectral types O7$-$O4 (\citealt{panagia73}). However, such a massive star has not been detected (\citealt{cruz-gonzalez07}). Moreover, the luminosity of this object would exceed the total luminosity of the DR 21 region of 10$^{5}$ L$_{\odot}$ (\citealt{harvey77}). Recently, \citet{zapata13} claimed that the outflow appears to have been produced by an explosive event.

In this paper we present the results of X-ray Chandra observations of the massive star-forming region DR 21. We complement these data with publicly available optical and near-IR and mid-IR observations. 
The paper is organized as follow. In Section \ref{data-analysis} we describe the data selection. We present the Chandra X-ray source catalog of the DR 21 cluster obtained from the Chandra XAssist Source List archive, and discuss the source membership to the cluster, possible contamination and completeness. We also present the optical and IR catalogs used to complement the X-ray survey. In Section \ref{results} we show the results of the cross-correlation between the Chandra catalog with the optical and IR catalogs, and study the spatial distribution, density, evolutionary stages and extinction of the stellar cluster. In Section \ref{star-formation} we discuss the implications of our results in the formation of the cluster and of the massive stars in the region. In view of the stellar population revealed by Chandra, in Section \ref{origin-outflow} we evaluate the possibility of a stellar merger as origin of the large-scale and highly-energetic outflow found in the DR 21 core. Finally, in Section \ref{summary} we summarize our findings and give our conclusions. 

\begin{figure}
\centering 
\includegraphics[angle=0,width=7.25cm]{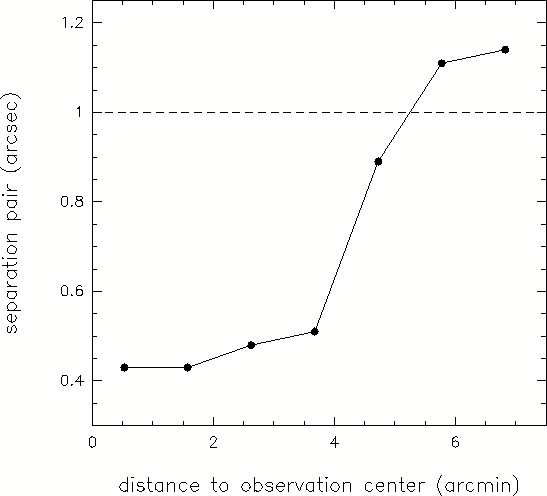}
\caption{Average position separation of pair of counterparts when crossing the 4 Chandra observing run samples two by two, as a function of the distance to the pointing center of the observation. The number of sources in each bin is 27, 41, 53, 11, 21 and 28, from lower to higher distances.}
\label{fig-separation}
\end{figure}

\begin{table}
\caption{Chandra observing runs in the DR 21 region.}            
\label{table-runs}      
\centering 
\begin{scriptsize}  
\begin{tabular}{c c c } 
\hline 
Observing run & Date           & Time exposure (ks) \\ 
\hline
07444 & 2007-08-22 & 48  \\
08598 & 2007-11-27 & 20 \\
09770 & 2007-11-29 & 19 \\
09771 & 2007-12-02 & 9 \\
\hline       
\end{tabular}
\end{scriptsize}         
\end{table}

\section{Data analysis}
\label{data-analysis}

In our analysis we used archival stellar catalogs in X-rays, optical and IR wavelengths. 
We also made use of the Spitzer 4.6 $\mu$m and 8 $\mu$m images (Cygnus-X Legacy Project), and the SCUBA 850 $\mu$m data (\citealt{matthews09}).

\subsection{Stellar catalogs}

\subsubsection{X-rays source catalog (Chandra)}
\label{chandra-catalog}

\begin{table*}
\caption{DR 21 Chandra source catalog, with SDSS, UKIDSS and Spitzer counterparts. Only the 15 first sources are shown here for form and content. The complete list is available in the on-line version of the paper as supplementary material.}            
\label{catalog} 
\tabcolsep 6pt     
\centering 
\begin{tabular}{c c c c c c c c} 
\hline 
DR21-X number & RA (J2000) & DEC (J2000) &  Counts & log $L_{\rm X}^{a}$ (erg s$^{-1}$)  & SDSS & UKIDSS & Spitzer-SSTCYGX  \\ 
\hline
  1	&      20  39  00.04	&      42  19 36.8	&      36	&   30.89       & - & J203900.04+421936.6 &  -   \\
  2	&      20  39  00.60	&      42  19 36.9	&       9	&   30.27 	    & - &  - &  -   \\
  3	&      20  39  00.35	&      42  19 31.0	&       9	&   30.28	 	& - & J203900.35+421931.7&   -   \\
  4	&      20  38  59.95	&      42  19 44.6	&       8	&   29.87	 	& - &  -&   -   \\
  5	&      20  39  00.92	&      42  19 41.4	&       8	&   30.22	 	& - & - &  -    \\
  6	&      20  38  59.55    &      42  19 33.3	&       9	&   30.61	 	& - &  - &  -   \\
  7	&      20  39  01.30	&      42  19 36.5	&      12	&   30.43	 	&  -&  - &  -   \\
  8	&      20  39  01.11	&      42  19 44.5	&      45	&   30.98	 	& - & J203901.11+421944.3&   -   \\
  9	&      20  38  59.09	&      42  19 39.5	&      12	&   30.03	 	& - & J203859.08+421939.5&   -  \\
 10	&      20  39  00.12	&      42  19 51.6	&      24	&   30.31	 	& - &  - &  -  \\
 11	&      20  38  59.60	&      42  19 53.4	&      13	&   30.45		& - &  J203859.61+421953.2  & J203859.65+421952.9 \\
 12	&      20  39  01.02	&      42  19 54.3	&      	5   &   30.10	 	& - & J203901.01+421954.4 &   \\
 13	&      20  39  00.45	&      42  19 59.0	&      20	&   30.63	 	& - &  - &  -  \\
 14	&      20  39  01.19	&      42  19 18.9	&       7	&   30.21	 	& - &  - &  -  \\
 15	&      20  39  01.21	&      42  19 18.6	&      84	&   30.85	  & - &  - &  -  \\
\hline       
\end{tabular}
\begin{list}{}{}
\item[$^{\mathrm{a}}$]{X-ray luminosity not corrected from extinction.}
\end{list}
\end{table*}

We used the catalog of X-ray sources presented in the Chandra XAssist Source List archive (CXOXASSIST\footnote{http://heasarc.gsfc.nasa.gov/W3Browse/chandra/cxoxassist.html}, \citealt{ptak03}). This database provides Chandra data that have automatically reduced for sources with sufficient counts. 
The DR 21 region was monitored in four observing runs between August and December 2007, with net exposure times between 48 and 9 ks (Table \ref{table-runs}), with the Advanced CCD Imaging Spectrometer (ACIS) onboard the Chandra X-ray Observatory. The ACIS-I array, consisting of four X-ray CCDs, covered a full region of 17.4$\arcmin\times$ 17.4$\arcmin$, which comprises the full region of DR 21. The telescope position of the field target was RA$_{J2000}$ = 20$^h$ 39$^m$ 0.70$^s$ and DEC$_{J2000}$ = 42$^\circ$ 18$\arcmin$ 56.8$\arcsec$. The Chandra point-spread function at the on-axis position is $\sim$0.5$\arcsec$. For source detection, the CXOXASSIST catalog used a wavelet transform detection algorithm implemented as the WAVDETECT program within the Chandra Interactive Analysis of Observations (CIAO) package version 4.3.0. with a threshold significance of 10$^{-6}$.
Due to intrinsic X-ray variability of PMS stars (expected to be the vast majority of the X-ray sample, see \citealt{getman05b}), not all sources are detected in the 4 observing runs. Therefore, to obtain a complete catalog, we have cross-correlated the samples from the 4 different runs. 
We have estimated the relative positional error of the sources crossing the runs two by two, and calculating the separation between pair of counterparts within separations from 0$\arcsec$ to 4$\arcsec$. In Fig. \ref{fig-separation} we plot the average separation as a function of the distance to the center of the observation. As expected, the positional error increases with the off-axis distance: remains below 1$\arcsec$ for distances $<$ 5$\arcmin$, and below 1.2$\arcsec$ in the outer parts of the region studied in this paper. Therefore, we have considered as the same source those sources from different runs that falls within 1$\arcsec$ for distances $<$ 5$\arcmin$ and within 1.5$\arcsec$ for distances $>$5$\arcmin$\footnote{We note that this cross-correlation provides properly the positions of the Chandra X-ray source catalog of the region, which is the interest for this work. However, for a rigorous derivation of the physical X-ray parameters of the sources (which is not the focus of this paper), a deeper image resulting for sum of the 4 observations runs would be more appropriate.}. In the case of sources observed in several runs, since some X-ray variability is present between epochs, we selected the observation with better signal-to-noise ratio\footnote{The values for signal-to-noise ratio have been determined in terms of the equivalent number of background fluctuations, defined as the net counts in the source over the square root of the background counts.}. 

\begin{figure}
\centering 
\includegraphics[angle=0,width=8.5cm]{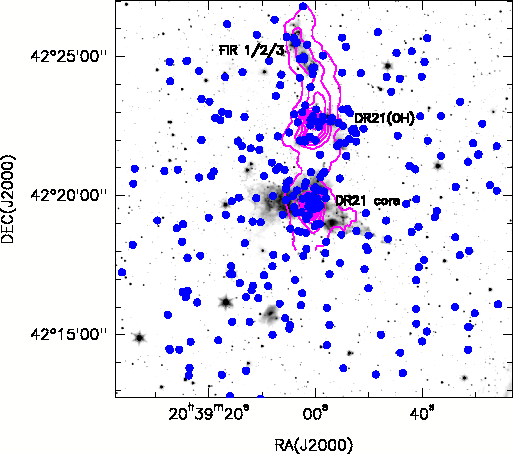}
   \caption{Spatial distribution of the X-ray Chandra catalog of the DR 21 cluster (blue dots). The background image is the 4.6 $\mu$m Spitzer image, and magenta contours correspond to dust emission detected by SCUBA at 850 $\mu$m (\citealt{matthews09}). The approximate size of the region shown is 6 pc $\times$ 6 pc.}
\label{fig1}
\end{figure}

In this paper we focus our analysis in a region with radius 7.5$\arcmin$ around the center of DR 21, obtaining a final catalog with 281 X-ray sources. The spatial distribution of the full X-ray sample is shown in Fig. \ref{fig1}. We present the list of sources in Table \ref{catalog}, with their coordinates, number of counts and luminosities (not corrected by extinction). The luminosities in the 0.3 to 8 keV energy range have been calculated by using $L_{\rm X}=4\pi$ $d^2$ $F_{\rm X}$, where $d$ is the distance to the region and $F_{\rm X}$ has been obtained from CXOXASSIST\footnote{The values for the X-ray luminosities presented in CXOXASSIST are calculated from a power-law model with a default slope of 1.8. Although little changes are expected, we note that for a more rigorous derivation of the luminosities of the stars (which is out of the scope of this paper), a two-temperatures plasma model would be more adequate (see \citealt{getman05a}).}.

\subsubsection{Optical source catalog (SDSS) and IR source catalogs (UKIDSS and Spitzer)}

\begin{figure}
   \centering 
\includegraphics[angle=0,width=8.5cm]{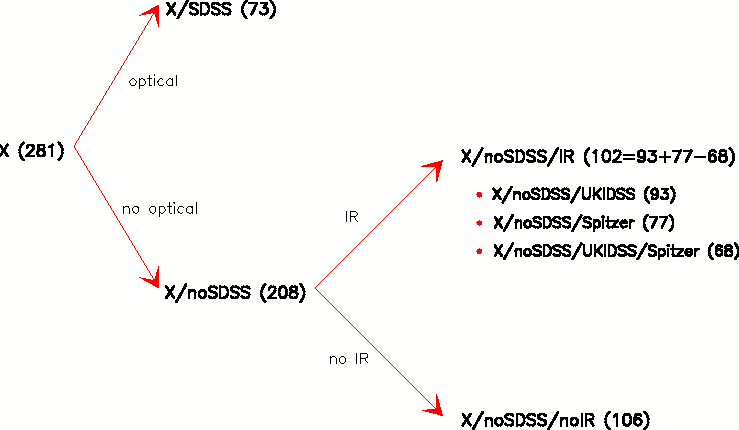}
   \caption{Scheme of the different samples and subsamples we used in this work, along with the number of sources in each group.}
\label{fig-esquema}
\end{figure}

We complemented the X-ray source catalog with the optical catalog provided by the Sloan Digital Sky Survey (SDSS, data release 9\footnote{http://www.sdss.org/}), and with two archival infrared (IR) surveys: i) the UKIDSS (UKIRT Infrared Deep Sky Survey, \citealt{lawrence07}) Galactic Plane Survey (GPS, data release 6, \citealt{lucas12}), and ii) Spitzer Space telescope catalog provided by the Cygnus-X Legacy survey \footnote{http://irsa.ipac.caltech.edu/cgi-bin/Gator/nph-scan?projshort=SPITZER}$^{,}$\footnote{http://www.cfa.harvard.edu/cygnusX/}.
The UKIDSS survey observed the region at $J$ (1.25 $\mu$m), $H$ (1.65 $\mu$m) and $K$ (2.2 $\mu$m) bands with the Wide Field Camera (WFCAM) on the United Kingdom Infrared Telescope (UKIRT), with typical 90$\%$ completeness limits in uncrowded fields of $K$=18.0, $H$=18.75 and $J$=19.5. The Infrared Array Camera (IRAC) onboard the Spitzer telescope observed in band 1 (3.6 $\mu$m), band 2 (4.5 $\mu$m), band 3 (5.8 $\mu$m) and band 4 (8.0 $\mu$m), with 90$\%$ completeness limits in uncrowded fields of 14.98, 14.87, 13.82, and 12.60, respectively (\citealt{beerer10}).
We cross-checked the DR 21 X-ray catalog with these catalogs, searching counterparts for the X-ray sources within a radius of 1$\arcsec$. In the case that two sources fall within this radius, we selected the best match. In Table \ref{catalog} we also present the association between X-ray and SDSS, UKIDSS and Spitzer sources. In Fig. \ref{fig-esquema} we show an scheme summarizing the subsamples resulting from the cross-correlation with the different catalogs.

\subsection{Foreground/background contamination}
\label{contamination}

While foreground and background galactic sources produce significant contamination in the IR studies of young clusters, X-ray surveys have the advantage of presenting very little galactic contamination, because PMS stars emit X-rays at levels 10-10$^{4}$ times higher than foreground/background main-sequence stars (\citealt{preibisch05a}). Several works (\citealt{getman05a,townsley11,mucciarelli11}) have shown that X-ray studies are very effective in revealing the embedded young stellar population in clusters, discriminating the PMS cluster members from unrelated older stars. 
Besides the sensitivity limit, which will be discussed in Section \ref{completeness}, this is mainly because IR surveys are affected by a much higher foreground/background contamination. However, X-ray sources could still be confused with extragalactic (EG) sources. In that case, other criteria, like spatial distribution, must be used to distinguish between cloud members and unrelated sources.

\begin{figure*}
   \centering 
\includegraphics[angle=0,width=9cm]{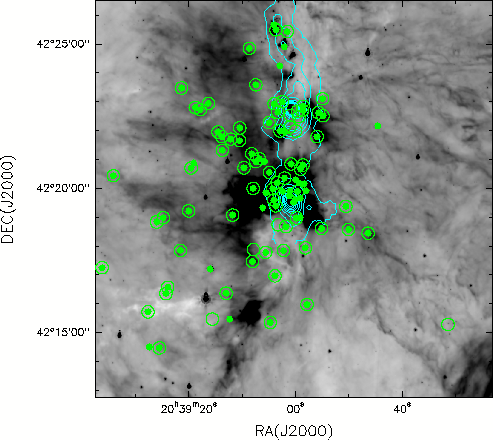}
\hspace*{0.5cm}
\includegraphics[angle=0,width=7.25cm]{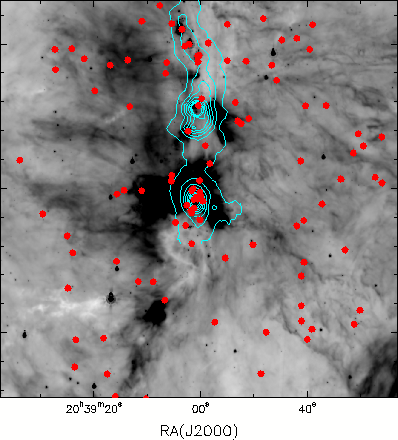}
   \caption{Spatial distribution of the X/noSDSS/UKIDSS source sample (green dots, left panel) and X/noSDSS/Spitzer source sample (open green circles, left panel), and X/noSDSS/noIR source sample (red dots, right panel). The background image is the 8.0 $\mu$m Spitzer image, which traces warm dust emission (\citealt{beerer10}). The light blue contours indicate cold dust emission detected by SCUBA at 850 $\mu$m (\citealt{matthews09}).}
\label{fig2}
\end{figure*}

To discriminate the embedded young stellar population from foreground stars, we cross-checked the X-ray source catalog with the SDSS optical survey. We found that 73 out of the 281 X-ray sources (26$\%$) have SDSS counterpart (X/SDSS sample). Most of these sources have visual extinctions $A_{\rm V}<$ 5 mag\footnote{We have estimated their extinction from their location in a $J-H$ vs. $H-K$ diagram (not showed), using the colors of their UKIDSS counterparts.}, confirming that they are likely galactic foreground stars unrelated with the cluster. This percentage is the same that was found by \citet{mucciarelli11} in the S255-IR cluster. The remaining 208 sources (X/noSDSS sample) are candidates to DR 21 cluster members. We cross-checked this non-optical X-ray sample with the UKIDSS and Spitzer source catalogs finding that 102 X-ray sources have UKIDSS and/or Spitzer counterparts (X/noSDSS/IR sample). The spatial distribution of these very likely cluster members is shown in left panel of Fig. \ref{fig2}.

The remaining X-ray sources without UKIDSS/Spitzer counterpart (106 sources, X/noSDSS/noIR sample) can be new cluster members which are not detected at IR wavelengths (likely because they suffer high extinction) or EG contamination, whose infrared emission is usually too faint to be detected in the IR catalogs. To estimate the expected number of EG sources we used the contamination detected in the more sensitive Chandra observations of the Orion Nebula Cluster (Chandra Orion Ultra Deep Project, COUP, \citealt{getman05a}, $\sim$850 ks) and the Carina Nebula (Chandra Carina Complex Project, CCCP, \citealt{townsley11}, $\sim$80 ks). The COUP analysis classified 10$\%$ of the total detected sources as EG contaminants. In DR 21 this would be $\sim$28 sources. The CCCP observation exhibits $\sim$1.5 EG contaminants per deg$^{2}$. Scaling these numbers to the field-of-view of DR 21, we obtain $\sim$20 sources. Taking into account that these observations are deeper, and then more EG contamination is expected (especially in the very deep COUP), we set an upper limit of  $\sim$20 EG contaminants in the X/noSDSS/noIR sample. Namely, more than 80$\%$ of these sources are likely new cluster members detected by Chandra for the first time. We can also use the spatial distribution as a criteria to discriminate between heavily embedded cluster members and background EG sources\footnote{The presence of X-ray flare events has been also used in previous works (see \citealt{getman05b}) to identify cluster members from EG contaminants. However, we can not use this criteria because X-ray light curves for the DR 21 sources are not available.}. The former are expected to be located preferentially in the more extincted region, which is the N-S cloud traced by the 850 $\mu$m SCUBA dust emission, while the latter are expected to be distributed throughout the whole field. The right panel of Fig. \ref{fig2} clearly shows that a significant fraction ($\sim$30$\%$) of the X/noSDDS/noIR sample are distributed along the cloud, peaking at the DR 21 core. We conclude thus that these sources are heavily obscured cluster members. The classification of the other sources ($\sim$80) is less clear. According to the estimates of the EG contamination, up to $\sim$20 could be contaminants, while the remaining would be stars of the cluster.
 
\begin{figure}
   \centering 
\includegraphics[angle=0,width=8cm]{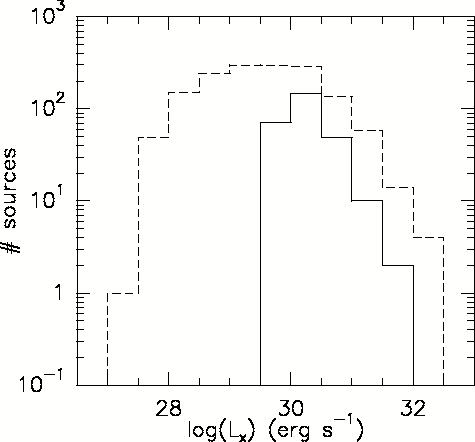}
\caption{Distribution of X-ray luminosities (not corrected from extinction) of the full Chandra sample, compared with that of the COUP sample (dashed line).}
\label{fig3}
\end{figure}

\subsection{Completeness}
\label{completeness}

We can estimate the sensitivity limits of the X-ray source catalog of DR 21 from the distribution of X-ray luminosities of the detected sources. In Fig. \ref{fig3} we show the histogram of the X-ray luminosities (not corrected from extinction) of the DR 21 population. The minimum value of log$L_{\rm X}$  is 29.69 erg s$^{-1}$. Using the correlation between $L_{\rm X}$ and the stellar mass found by \citet{preibisch05a} in the Orion Nebula Cluster, this is equivalent to a lower mass limit of $\sim$0.38 M$_{\odot}$. Considering a typical initial mass function (IMF) of young clusters (\citealt{chabrier03}), and a range of masses in the cluster between 0.02 and 10 $M_{\odot}$, the population detected would represent $\sim$52$\%$ of the total cluster population.

An alternative method to estimate the fraction of stars that Chandra observations has detected towards DR 21 is to calculate the fraction of stars that COUP detected  with luminosities above the DR 21 limit of log $L_{\rm X}$ = 29.69 erg s$^{-1}$. Fig. \ref{fig3} shows the histogram of X-ray luminosities (also without extinction correction) of the Orion Nebula Cluster (COUP, \citealt{getman05a}), whose census can be considered nearly complete\footnote{With the exception of the population heavily embedded in the Orion Molecular Cloud, see \citet{grosso05}.}. It is clear that the Chandra DR 21 observation is biased to detect the  brighter X-ray sources. The fraction of COUP sources with log$L_{\rm X}>$ 29.69 erg s$^{-1}$ is 45$\%$. This is a value similar to the one obtained previously, indicating that the Chandra observation of DR 21 detected around half of the total cluster population in those region in the absence of extinction. 
 
This fraction is obviously even lower in the obscured regions, like the N-S cloud traced by the 850 $\mu$m dust emission. \citet{grosso05} estimated that in the obscured regions in Orion like the Orion Hot Core and OMC1-S regions (which suffers $A_{\rm V}>$ 25 mag) the X-ray observation detected $\sim$48-63$\%$ of the sources due to obscuration. Therefore, in regions with such high extinctions, like the DR 21 core (see Section \ref{UKIDSS}), it is expected that the X-ray observations only detect a half of the population due to extinction. 
Furthermore, we need to take into account the likely presence of unresolved binaries that the spatial resolution of Chandra is unable to resolve. 

To summarize, the Chandra observations of DR 21 have detected a new population of deeply embedded stars. Our analysis indicates that we underestimate the population of stars by a factor of 2 for the regions with low extinction and by a factor of $>$4 for the highly extincted regions ($A_{\rm V}>$ 25 mag.).

\section{Results: The new X-ray PMS stellar population in DR 21}
\label{results}

\subsection{X-ray sources with UKIDSS counterparts}
\label{UKIDSS}

The cross-correlation with the UKIDSS catalog shows that 162 X-ray sources within the sample have UKIDSS counterpart (58$\%$). For comparison, in the S255-IR star-forming region, the embedded cluster observed with Chandra (\citealt{mucciarelli11}) has a similar IR counterpart fraction of 63$\%$. Considering the X/noSDSS sample, 93 of the 208 sources (45$\%$) have UKIDSS counterparts (X/noSDSS/UKIDSS sample). The spatial distribution of this sample is shown in the left panel of Fig. \ref{fig2}. 
It shows a centrally peaked distribution of the stars, with the DR 21 core as the main feature. The morphology of this population shows a filament-like feature towards the NE, with origin in the DR 21 core. In Section \ref{star-formation} we will discuss the possible implications for star formation inferred from this observed structure.

\begin{figure}
   \centering 
\includegraphics[angle=0,width=8.0cm]{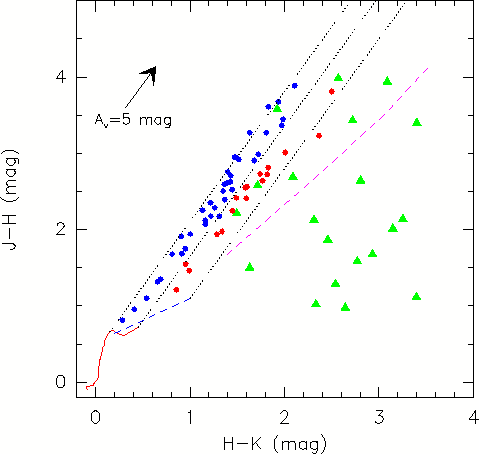}
   \caption{$J-H$ vs. $H-K$ color-color diagram of the X/noSDSS/UKIDSS source sample. Stars with/without near IR excesses are denoted by blue/red dots, respectively. The green triangles are lower limits for those sources detected in $H$ and $K$ bands, but not in the $J$ band. The solid red line indicates the IR colors of ZAMS stars with masses from 0.1 to 7 M$_{\odot}$ from the theoretical evolutionary tracks of \citet{siess00}. The dashed blue line denotes the unreddened locus of intrinsic near-IR excesses due to presence of disks from (\citealt{meyer97}). The black dotted lines are reddening vector using the extinction law from \citet{rieke85}, extended from the ZAMS colors and the from unreddened locus. The magenta line denotes the minimum value of $J-H$ that stars with masses 0.1$-$4 M$_{\odot}$ and an age of 1 Myr (approximated mean age for the cluster, \citealt{beerer10}) must have to be detected in the $H$ band and not in the $J$ band.}
\label{fig5}
\end{figure}

In Fig. \ref{fig5} we show the $J-H$ vs. $H-K$ color-color diagram of the X/noSDSS/UKIDSS sample. 
Stars without strong emission from a disk have colors typical of extincted zero-age main sequence (ZAMS) stars and populate the confined region between the two leftmost black dotted lines. Stars with the presence of accretion disks, populate the region between the two rightmost dotted lines, while those with additional IR excess dominated by infalling envelopes should be found in the rightmost region of the diagram. 
Besides the sources detected in $J$, $H$ and $K$ bands, there is a significant fraction of X/noSDSS/UKIDSS stars detected in $H$ and $K$ bands, but not in the $J$ band due to extinction. For those sources, we have considered lower limits for the color $J-H$, using the UKIDSS detection limit for the J band ($m_{\rm J}$=19.5 mag, \citealt{lawrence07,lucas12}). 

\begin{figure*}
\centering 
\includegraphics[angle=0,width=8cm]{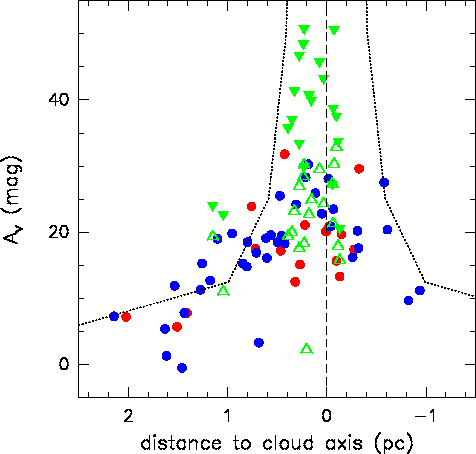}
\hspace*{1.5cm}
\includegraphics[angle=0,width=8cm]{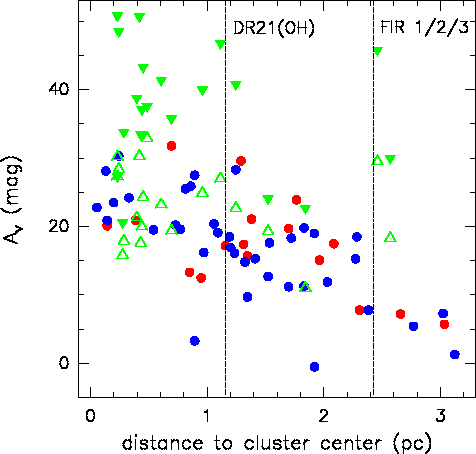}
\caption{{\it Left panel:} Values of the extinction of the X/noSDSS/UKIDSS sample derived from the $H-K$ colors of the stellar population versus the distance to the approximate axis of the cloud detected at 850 $\mu$m by \citet{matthews09} (vertical dashed line). Blue and red dots correspond to stars without/with near IR excesses, respectively. Open and filled green triangles indicate the lower and upper limits, respectively, for the stars not detected in the $J$ band. The dotted curve indicates the column density profile obtained by \citet{hennemann12} using $Herschel$ data. Distances to the east are defined as positive. {\it Right panel:} Values of the extinction towards the stellar population versus the distance to the cluster center, located in the DR 21 core. The colors and symbols are the same as in the left panel. The dotted vertical lines indicate the location of the DR 21(OH) and FIR 1/2/3 regions.}
\label{fig6}
\end{figure*}

\begin{figure}
   \centering 
\includegraphics[angle=0,width=8.5cm]{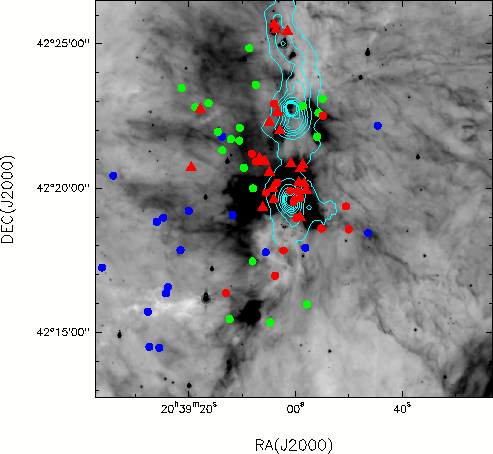}
   \caption{Spatial distribution of the X/noSDSS/UKIDSS source sample as a function of extinction: stars with $A_{\rm V}<$15 mag (blue dots), stars with 15 mag  $<A_{\rm V}<$ 20 mag (green dots), and stars with  $A_{\rm V}>$20 mag (red dots). We also include (red triangles) the sources not detected in the $J$ band, expected to be more embedded (see Fig. \ref{fig6}). The background image is the 8.0 $\mu$m Spitzer image and the light blue contours indicate the emission detected by SCUBA at 850 $\mu$m (\citealt{matthews09}).}
\label{fig7}
\end{figure}

From the color-color diagram we estimate the extinction suffered by the stars.\footnote{The extinction could also be derived from the values for the hydrogen column density $N_{\rm H}$ from the fitting of the X-ray spectra of the sources. However, CXOXASSIST only provides $N_{\rm H}$ values for sources with $>$ 100 counts (less than 10 sources). A more detailed spectroscopy study of the X-ray spectra for the full sample would be needed to calculate $N_{\rm H}$.}   
Using the extinction law by \citet{rieke85}, the color index (in absence of significant near IR emission from disk or envelopes) is related with the visual absorption $A_{\rm V}$ with the expression: $A_{\rm V}$= ($H-K$-0.2)/0.063\footnote{We adopted $H-K$=0.2 as the typical intrinsic color of most stars (\citealt{siess00}).}. We used this expression to calculate the extinction of sources without near IR excesses. According to Fig. \ref{fig5} the presence circumstellar disks produce an average excess in the $H-K$ color of $\sim$0.3 mag. We estimate then the extinction suffered for the stars with excesses subtracting this contribution to the measured value of $H-K$. For those sources not detected in $J$ we calculate lower an upper limits for the extinction.
For the upper limits we consider that the star do not show a near IR excess, i.e., the $H-K$ is only due to extinction (without contribution from disk/envelope). For the lower limits we consider a maximum disk/envelope contribution, assuming that the $J-H$ values are those corresponding to the limit for detection in $J$ band (magenta dashed line in Fig. \ref{fig6}), and subtracting the $H-K$ excess due to disk/envelope material. 
Following \citet{rivilla13a}, we show in Fig. \ref{fig7} the spatial distribution of the X/noSDSS/UKIDSS sources as a function of extinction: stars with $A_{\rm V}<$ 15 mag (blue dots), stars with 15 mag $<A_{\rm V}<$ 20 mag (green dots), and stars with $A_{\rm V}>$ 20 mag (red dots). We also include (red triangles) the sources not detected in the $J$ band, expected to be highly embedded. This analysis shows that the X/noSDSS/UKIDSS sources with higher extinction follow the N-S cloud, those with intermediate extinction are concentrated along the NE stellar filament, and those with lower extinction are more distributed and located at larger distances from the cloud. The sources not detected in $J$ are concentrated along the cloud, mainly surrounding the DR 21 core, DR 21(OH) and the FIR 3 source, confirming that they are more embedded objects.

In Fig. \ref{fig6} (left panel) we show the extinction versus the distance to a N-S axis following the 850 $\mu$m emission from the cloud. It is clear that the extinction increases when the distance to the cloud axis decreases. The X/noSDSS/UKIDSS sources detected in all three near IR bands show extinctions up to $A_{\rm V}$=30 mag, peaking in the inner part. This extinction peak is similar to the one found in the low resolution extinction map from \citet{schneider06} from the Two Micron All Sky Survey (2MASS). However, the $Herschel$ observations from \citet{hennemann12} showed that the extinction is even higher in the inner region of the cloud (dashed black curve in Fig. \ref{fig6})\footnote{We have used the relation $N_{\rm H}$/$A_{\rm V}$=2$\times$10$^{21}$, which is an intermediate value between the relations found by \citet{ryter96} and \citet{vuong03}.}. Since the UKIDSS survey is deeper than the 2MASS survey, we expect to detect stars more embedded in the cloud. Actually, as indicated Fig. \ref{fig6}, the X/noSDSS/UKIDSS sources not detected in $J$ band are located in the inner region of the cloud, and show extinctions between $A_{\rm V}$=20 mag and $A_{\rm V}$=50 mag, consistently with the column density profile from \citealt{hennemann12}. In Section \ref{cumulative} we will show that in the very inner region there are even more extincted stars only detected by Chandra.

In the right panel of Fig. \ref{fig6} we show the value of the extinction with respect to the center of the cluster, located in the DR 21 core. With the only exception of some high-extincted stars around  DR 21(OH) ($\sim$1 pc) and the FIR 1/2/3 region ($\sim$2.2 pc), there is also a clear overall trend in the extinction, gradually decreasing with increasing distance to the cluster center.

\begin{figure*}
   \centering 
\includegraphics[angle=0,width=7.5cm]{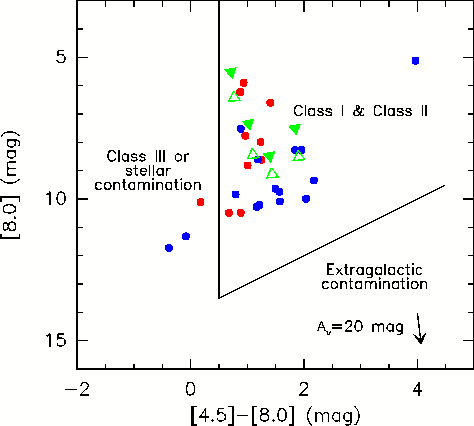}
\hspace*{0.6cm}
\includegraphics[angle=0,width=7.5cm]{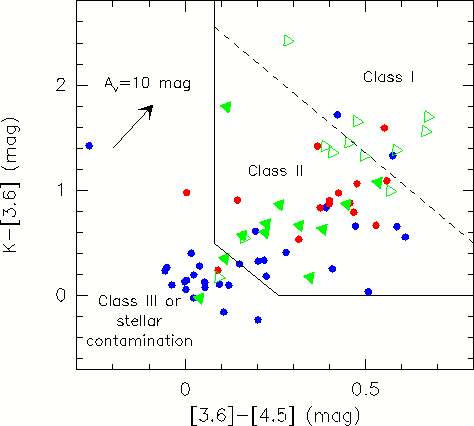}
   \caption{{\it Left panel:} Extinction-corrected magnitude-color diagram of the X/noSDSS/UKIDSS/Spitzer sample source sample. Blue and red dots correspond to stars without/with near IR excesses, respectively. Open and filled green triangles indicate the fluxes corrected with lower and upper limits of the extinction, respectively, for the stars not detected in the $J$ band. The regions corresponding to different classes of objects are indicated (\citealt{jorgensen06}), along with and extinction vector from the \citet{flaherty07} extinction law. {\it Right panel:} Extinction-corrected color-color diagram of the Chandra/Spitzer source sample. The symbols are the same as in the left panel. The regions correspond to the different classes and the extinction vector are indicated.}
\label{fig8}
\end{figure*}

\subsection{X-ray sources with Spitzer counterparts}
\label{spitzer}

The cross-correlation between the Chandra and the Spitzer-IRAC source catalogs shows that 145 
 X-ray sources have UKIDSS counterparts (52$\%$). Considering only the X/noSDSS sample, 77 of the 208 
 sources (37$\%$) have Spitzer counterparts (X/noSDSS/Spitzer sample), of which 68 have UKIDSS counterpart (X/noSDSS/UKIDSS/Spitzer sample). Many works have shown that the IRAC data are very efficient identifying young stellar objects, because it detects excess emission well above that expected from reddened stellar photospheres that originates from the dusty circumstellar disks and envelopes surrounding young stars (\citealt{allen04,harvey06,jorgensen06}).

In the right panel of Fig. \ref{fig1} we presented the spatial distribution of the X/noSDSS/Spitzer sample. Similarly to the X/noSDSS/UKIDSS sample, the morphology of this population shows the NE filament-like feature. Compared with UKIDSS, Spitzer detected less sources in the inner DR 21 core. Besides the higher extinction in the DR 21 central region (which also affect UKIDSS), this is likely due to two other factors: i) the lower Spitzer spatial resolution prevents a complete census of crowded dense clusters (\citealt{mucciarelli11}) like the one found in the DR 21 core; ii) the extended emission from the large outflow (detected in all 4 IRAC bands) decreases the sensitivity to detect point-like stellar sources\footnote{The decrease of sensitivity due to extended emission from the large outflow can also affect the UKIDSS observations in the $K$ band, but barely in the $H$ and $J$ bands, where the outflow is not detected.}.

From the IRAC colors, it is possible to classify the stars as Class I, Class II and Class III (according to the classification from \citealt{wilking83}), which indicates their evolutionary stage, from earlier to more evolved.
Since extragalactic sources may also be misidentified as young stars or protostars, we use a magnitude-color diagram ([8.0] vs. [4.5]-[8.0], \citealt{whitney03}, \citealt{jorgensen06}, \citealt{harvey06}) that allows the discrimination of EG contamination. 
The left panel of Fig. \ref{fig8} shows the extinction-corrected [8.0] vs. [4.5]-[8.0] diagram of the X/noSDSS/Spitzer sample. 
It is remarkable that none of the sources fall in the region expected for extragalactic contaminants, in agreement with the low extragalactic contamination discussed in Section \ref{contamination}.

Although the [8.0] vs. [4.5]-[8.0] diagram effectively discriminate EG contaminants, the relatively poor sensitivity and spatial resolution of the longer wavelength IRAC 8.0 $\mu$m band (see \citealt{lucas12}) prevents the classification of many stars. For this reason, to classify the sources in their evolutionary stages we use the $K$-[3.6] vs [3.6]-[4.5] diagram, which in the DR 21 region is able to classify a number of stars twice than the [8.0] vs. [4.5]-[8.0] diagram.
The right panel of Fig. \ref{fig8} shows the $K$-[3.6] vs [3.6]-[4.5] extinction-corrected diagram of the X/noSDSS/Spitzer sample.
Since background planetary nebulae and asymptotic giant branch (AGB) stars may also be misidentified as young stars, some of the stars in the Class III region could be stellar contamination unrelated with the cluster. However, the cross-correlation between the Spitzer and Chandra samples makes unlikely this possibility, because X-ray emission from PMS stars is significantly higher than that from main sequence or post-main sequence stars.

The diagram shows that most of the stars with near IR excesses (Section \ref{UKIDSS}) are Class II objects, while those without excesses are Class III objects. This is expected because Class II objects are younger PMS stars that still exhibit optically thick circumstellar disks, while Class III objects are more evolved PMS stars with colors more similar to naked stellar photospheres\footnote{Although we note that it is well known that longer IR wavelengths ($L$-band or Spitzer bands) are much more effective (and hence more reliable) to detect excesses from the disk and envelopes (\citealt{lada00}).}. 

\begin{figure}
   \centering 
\includegraphics[angle=0,width=8.5cm]{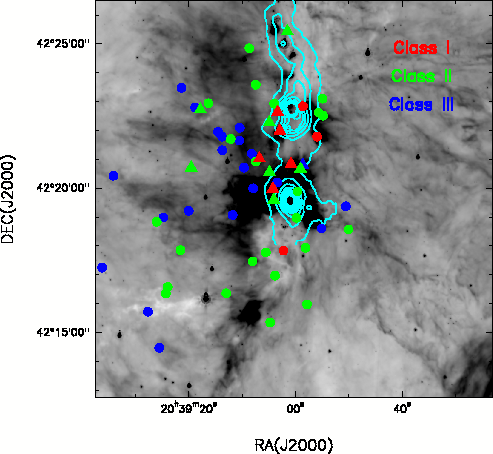}
   \caption{Spatial distribution of the X/noSDSS/UKIDSS/Spitzer sample, classified by evolutionary PMS star stages using the Spitzer K-[3.6] vs. [3.6]-[4.5] diagram. The dots denotes stars detected in all 3 near IR bands ($JHK$) and triangles stars not detected in $J$ band. Different colors correspond to different evolutionary stages: Class I (red), Class II (green), and Class III (blue). We consider Class I candidates (red triangles) those stars not detected in $J$ band whose flux corrected with the lower extinction fall in the Class I region of the color-color diagram. The background image is the 8.0 $\mu$m Spitzer image and the light blue contours indicate the emission detected by SCUBA at 850 $\mu$m (\citealt{matthews09}). }
\label{fig9}
\end{figure}

The spatial distribution of the different evolutionary classes from the $K$-[3.6] vs [3.6]-[4.5] diagram is shown in Fig. \ref{fig9}. The youngest objects (Class I\footnote{We have included as Class I candidates those stars not detected in the $J$ band whose whose flux corrected with the lower extinction fall in the Class I region.}) are found along the N-S cloud, while the subsequent evolutionary stages are more distributed. The NE stellar filament is composed by mainly Class II/III stars, suggesting that this stellar population is more evolved than that embedded in the cloud.

\begin{figure}
   \centering 
\includegraphics[angle=0,width=8cm]{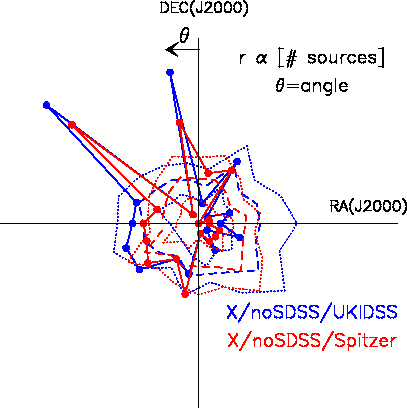}
   \caption{Angular distribution of the X/noSDSS/UKIDSS (solid blue line) and X/noSDSS/Spitzer (solid red line) stellar populations. The angle $\theta$ runs East from North as indicated. The radial distance to the center is proportional to the number of sources detected in angular bins of 20 deg. The dashed and dotted lines indicate average and $\pm$1$\sigma$ standard deviation value for each angular bin, respectively, considering randomly distributed stellar distributions from 10 Montecarlo simulations.}
\label{fig12}
\end{figure}

\subsection{Angular distribution of the stellar population: the NE filament}
\label{angular}

We showed in Section \ref{results} that the X/noSDSS/UKIDSS and X/noSDSS/Spitzer samples exhibit in DR 21 an elongated structure from the DR 21 core towards the NE that resemble these radial filaments. In Fig. \ref{fig12} we plot the angular distribution of the two stellar samples, in angular bins of 20 deg. The presence of the NE stellar filament is evident at $\sim$50 degrees. There is also other peak in the angular distribution, at $\sim$10 degrees, caused by the sub-clusters around DR21(OH) and FIR 1/2/3 (see Section \ref{stellar-densities}) rather than by an stellar filament. To confirm that the NE feature is real and not caused by a random fluctuation in the stellar spatial distribution, we carried out a Montecarlo analysis. For each sample, we generate artificial clusters with random distributions. We show in Fig. \ref{fig12} the average value and 1$\sigma$ standard deviation values of the set of 10 Montecarlo runs for each angular bin. While the simulations show an approximately isotropic distribution, the observed distribution in DR 21 clearly peaks in the two directions previously noted. Therefore, it is clear that the NE filament is well above the noise level, clearly showing that it is not due to a chance alignment of sources.

\begin{figure*}
   \centering 
\includegraphics[angle=0,width=12cm]{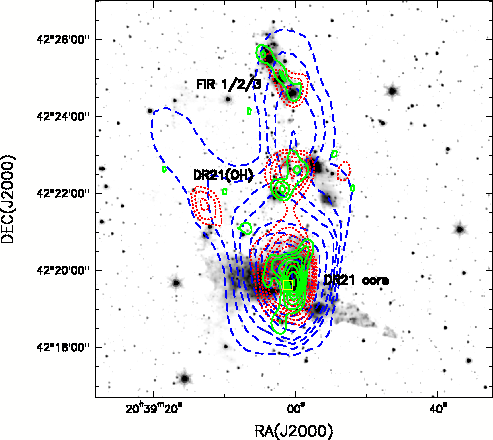}
\hspace*{0.5cm}
\includegraphics[angle=0,width=4.575cm]{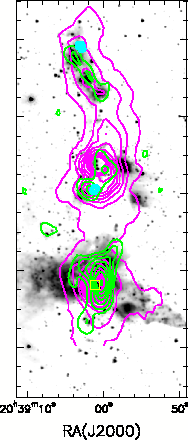}
   \caption{{\it Left panel}: Stellar density map of the X/noSDSS sample, computed  with the 120$\arcsec\times$ 120$\arcsec$ grid (0.87 pc $\times$ 0.87 pc, blue dashed contours), the 60$\arcsec\times$ 60$\arcsec$ grid (0.44 pc $\times$ 0.44 pc, red dotted contours) and the 30$\arcsec\times$ 30$\arcsec$ grid (0.22 pc $\times$ 0.22 pc, green solid contours). The contours are 6, 8, 10, 12, 17, 22, 27, 32, and 37 stars cell$^{-1}$ for the 120$\arcsec$ grid; 4, 5, 6, 7, 8, 11, 14 and 17 stars cell$^{-1}$ for the 60$\arcsec$ grid; and 2, 3, 4, 5, 6, 7, and 8 stars cell$^{-1}$ for the 30$\arcsec$ grid. The background image is the 4.6 $\mu$m Spitzer image. The positions of the massive star-forming regions (DR 21 core, DR 21(OH) and FIR 1/2/3) are labeled. The open yellow square indicates the position of the explosion center proposed by \citealt{zapata13} to explain the large CO outflow (seen in the 4.6 $\mu$m Spitzer image). {\it Right panel:} Zoom-in view of the N-S cloud, comparing the stellar density obtained with the 30$\arcsec\times$30$\arcsec$ grid (green contours) and the dust emission detected by SCUBA at 850 $\mu$m (magenta contours). The light blue circles indicate the positions of millimeter sources detected by \citet{bontemps10} that are candidates to form massive stars.}
\label{fig10}
\end{figure*}


\subsection{Stellar densities: sub-clusters of PMS stars}
\label{stellar-densities}

\subsubsection{Stellar density maps}
\label{stellar-density-map}

We have seen that the distribution of the Chandra sample (Fig. \ref{fig1}) are not homogeneously distributed over the DR 21 region, but centrally clustered towards the DR 21 core. 
Following \citet{rivilla13a}, we compute the stellar density of the X/noSDSS sample (208 sources) using a spatial gridding method. We count the number of stars in square cells with three different sizes: 120$\arcsec\times$120$\arcsec$ (0.87 pc$\times$0.87 pc), 60$\arcsec\times$60$\arcsec$ (0.44 pc$\times$0.44 pc) and 30$\arcsec\times$30$\arcsec$ (0.22 pc$\times$0.22 pc). We choose the cell sizes to match the size range of the clumps resulting from the fragmentation of molecular clouds (\citealt{williams00,saito07}) where massive star formation is expected to occur.
The results are presented in Fig. \ref{fig10}. We remark that, according to our completeness estimates, almost $\sim$ 50$\%$ of the cluster members still remain undetected in the regions without extinction due to sensitivity limits and only $<$ 25$\%$ of the sources are detected in the regions affected by extinction (Section \ref{completeness}). This means that the stellar densities that we obtained should be multiplied by factors of 2 or $>$ 4, respectively. Moreover, even these completeness corrected densities must be considered as lower limits due to the likely presence of binary and multiple systems not resolved by Chandra spatial resolution.

The lower resolution spatial gridding (120$\arcsec\times$120$\arcsec$ cell) reveals the general large-scale structures of the cluster. The NE stellar filament is also evident in the stellar density contours. The overall elongated N-S morphology of the X-ray population, peaking in the DR 21 core, agrees with the distribution of the molecular cloud traced by its dust emission at 850 $\mu$m measured with SCUBA. 

The spatial gridding with the 60$\arcsec\times$60$\arcsec$ cell reveals the more compact clustering of stars, with three well defined peaks along the cloud. The main peak is located in the central core of DR 21, coincident with the location of the young massive stars exciting the UC HII regions (\citealt{cyganowski03}). A secondary density peak is located at the molecular hot core DR 21(OH) (\citealt{chandler93}), an a third peak is located around the FIR 1/2/3 young massive stars.  

The morphology of the stellar density from the 30$\arcsec\times$30$\arcsec$ grid also peaks clearly in the three massive star forming regions, following the N-S cloud seen at 850 $\mu$m (see right panel in Fig. \ref{fig10}). 
With this higher resolution grid the DR 21(OH) population splits into two sub-clusters, with the more dense located SE the DR 21(OH) main peak. This sub-cluster is coincident with a secondary peak in the 850 $\mu$m map (also detected at 1200 $\mu$m, see \citealt{chandler93}), and located close of the center of the massive ($>$8 M$_{\odot}$) dense cores CygX-N48 MM1 and MM2 (\citealt{bontemps10}), that are candidates to form massive stars. The stellar sub-cluster revealed by Chandra is embedded in the cloud, with $A_{\rm V}>$ 20 mag\footnote{One of their members is detected in $H$ and $K$ bands, but not in $J$ band, show extinction $A_{\rm V}>$ 20 mag (Section \ref{UKIDSS}) and it is classified as Class I candidate (section \ref{spitzer}). Two other members are only detected in the $K$ band, and other is not detected by UKIDSS, suggesting that they are all more embedded in the cloud.}, and may represent the outer part of the more obscured CygX-N48 cluster of cores (log$N_{\rm H}\sim$23 cm$^{-2}$, from \citet{bontemps10}, i.e., $A_{\rm V}\sim$50 mag). Similarly, the massive dense cores CygX-N53 MM1 and MM2 are located very near the FIR 3 stellar density peak, suggesting that they could be related.

The DR 21 core shows a clear centrally peaked distribution. To provide quantitatively a value for the cluster size we use the {\it clustering parameter } ($\alpha$)  defined by \citet{rivilla13a} as the ratio between the number of sources found in the 0.22 pc$\times$0.22 pc cell and the number found in the 0.44 pc $\times$ 0.44 pc cell. 
Our results show that the clustering parameter in the DR 21 core of $\alpha$=0.47. If we assume a Gaussian distribution for the stellar density $\rho_{*}$ $\propto$ e$^{-(2r/\gamma)^{2}}$ (where {\em r} is the distance to the center of the cluster and $\gamma$ is defined as the cluster diameter), $\alpha$=0.47 corresponds to $\gamma$ $\sim$ 0.6 pc. This cluster size is in the typical range for molecular clumps $-$ regions of enhanced density within a molecular cloud $-$ which will typically form stellar clusters (\citealt{williams00,smith09b}).

\begin{figure}
   \centering 
\includegraphics[angle=0,width=8cm]{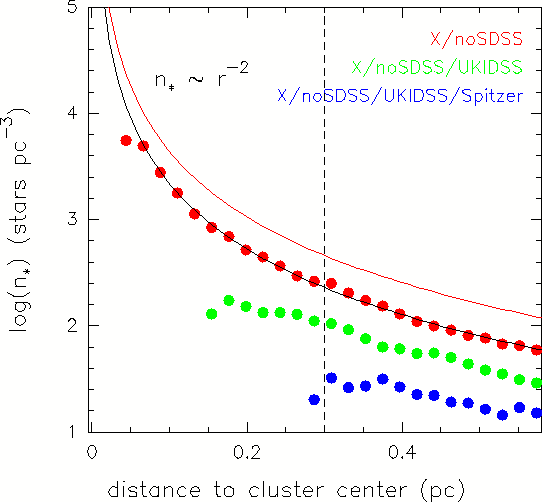}
   \caption{Cumulative stellar density of the X/noSDSS sample (red dots), compared with the X/noSDSS/UKIDSS (green dots) and  X/noSDSS/UKIDSS/Spitzer (blue dots) samples. We considered as inner distance that one at which at least 2 sources were detected. The fit of the only Chandra sample points follows a $r^{-2}$ profile (black curve). The red curve corresponds to the stellar density profile corrected by a factor of 2, due to the incompleteness of the Chandra observation (see Section \ref{completeness}). The dashed vertical line indicate the radius of the DR 21 core cluster calculated in section \ref{stellar-density-map}.}
\label{fig11}
\end{figure}

\subsubsection{Cumulative stellar density radial profile}
\label{cumulative}

We have shown in previous sections that the DR 21 low-mass stellar cluster is centrally condensed, with its density peak towards the DR 21 core. In this section we derive the radial stellar density profile of the DR 21 core cluster by calculating the cumulative stellar density with respect to the cluster center. We considered that the center of the cluster coincides with the location of the stellar density peak seen in the 0.22 pc $\times$ 0.22 pc grid (see Fig. \ref{fig10}). We count the number of stars of the X/noSDSS sample within concentric circles with radius increasing in 0.01 pc steps. We considered as the innermost concentric circle the one containing at least 2 stars. The results are presented in Fig. \ref{fig11}. According with the completeness discussion in Section \ref{completeness} we also present the stellar densities corrected by a factor of 2, which accounts for the limited sensitivity of the DR 21 Chandra observation. The cluster follows an approximate radial profile $\sim r^{-2}$ with the exception of the very inner region.
This is likely due to the higher extinction in the central part of the core, that prevents the detection of stars with weaker X-ray emission. 

We compare the cumulative radial density profile of the X/noSDSS sample with the X/noSDSS/UKIDSS and X/noSDSS/UKIDSS/Spitzer samples. Obviously the stellar density of the X/IR samples are lower because they are subsamples of the Chandra sample. The plot shows that Spitzer is not effective detecting cluster members in the inner $<$ 0.25 pc, because as already mentioned, it likely suffers from source crowding and from extended emission from the large outflow. The X/noSDSS/UKIDSS sample, less affected by extended emission, detect sources until 0.15 pc, but only Chandra has revealed the stellar population for smaller distances. The high extinction in the very inner region of the core (Section \ref{UKIDSS}) prevents the detection of the stellar population at IR wavelengths. However, the X-ray observations are capable to detect sources more deeply embedded (\citealt{mucciarelli11,rivilla13a}). Note that even the X-ray population seems to be heavily affected in the very inner region of the cluster, where the extinction is extremely high, and where the measured stellar density in the Chandra sample also deviates from the $r^{-2}$ profile.

The different profiles of the samples also reflect the centrally peaked structure of the extinction. The  profiles gradually deviate one from another for smaller distances to the cluster center. While the density of the X/noSDSS sample increases following the $\sim r^{-2}$ profile, the densities of the X/noSDSS/UKIDSS and X/noSDSS//UKIDSS/Spitzer exhibit less steep profiles. We interpret this as a consequence of the extinction in the cluster (Fig. \ref{fig6}).

\section{Competitive accretion scenario for cluster and massive star formation}
\label{star-formation}

One of the main results of this paper is that the density of PMS low-mass stars peaks at the massive star cradles: DR 21 core, DR 21(OH) and FIR/1/2/3 region. \citet{beerer10} also found that young stars tend to cluster around massive B stars in the Cygnus X region from their Spitzer data. This close association between the low-mass star clusters and the massive star cradles was also reported towards the ONC and OMC (\citealt{grosso05,rivilla13a}). This suggests that not only dense gas plays a role to form massive objects, but also a cluster of low-mass stars, in agreement with the "competitive accretion" theory of massive star formation (\citealt{bonnell04,bonnell06,smith09b}). According to this theory stars located in the densest regions of a stellar cluster benefit from the local potential wells created by the stellar population, and can gather enough mass to become massive. 

This scenario predicts three main features: i) massive stars are born in the low-mass stellar density peaks; ii) an overall centrally condensed structure of the extinction, because the potential well funnels gas and dust towards the cluster center; and iii) an overall mass segregation, with the more massive stars expected to be found in the central regions of the cluster.

\begin{figure}
   \centering 
\includegraphics[angle=0,width=8cm]{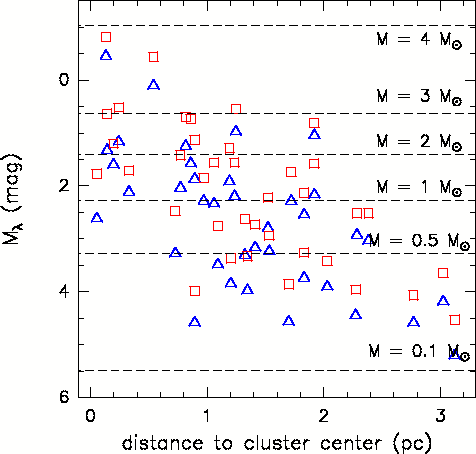}
   \caption{Infrared absolute magnitudes $M_{\lambda}$ of the X/noSDSS/UKIDSS sample without near IR excesses versus the distance to the cluster center. The blue triangles correspond to the $J$ filter and the red squares to the $H$ filter. The extinction correction was applied using the values of $A_{\rm V}$ obtained from the color index $H-K$, which is a good estimation of the absorption (see Section \ref{UKIDSS}). The horizontal dashed lines indicate the $J$ absolute magnitudes for 1 Myr stars for different masses, from the \citet{siess00} PMS models.}
\label{fig13}
\end{figure}

These 3 conditions are well fulfilled in the ONC/OMC region (\citealt{hillenbrand98,allison09a,rivilla13a}). As we have shown in Sections \ref{UKIDSS} and \ref{stellar-densities}, the first and second conditions are also met in DR 21 region. With the aim of studying the third condition, i.e., the level of mass segregation in the DR 21 region, we use the IR fluxes corrected by extinction ($M_{\lambda}$) of the X/noSDSS/UKIDSS sample without IR excesses. Whether the IR emission of the sources is not affected by the presence of disks or envelopes, the value of the corrected IR emission is a good proxy for the stellar masses (\citealt{siess00}). Fig. \ref{fig13} shows that the fluxes in the $J$ and $H$ bands gradually decrease with increasing distance to the cluster center. Obviously, the central structure of the extinction (right panel of Fig. \ref{fig6}) bias the detection of fainter objects in the inner region where the extinction is higher. However, the important point here is that there is not a population of bright (more massive) stars at larger distances where the extinction is expected to be lower. This pattern is very similar to the result obtained from the competitive accretion simulations from \citet{bonnell01} (see their Fig. 3), and shows that the mass segregation in the DR 21 complex could be explained by the competitive accretion scenario.

Supporting this idea, several works (\citealt{kumar07}, \citealt{schneider10}, \citealt{hennemann12}) have found several dusty and molecular filaments falling into the DR 21 N-S cloud. \citet{hennemann12} suggested that the mass accretion through these filaments is driven by the gravitational potential of the full cloud. These infalling filaments provide a continuous mass inflow into the cloud and replenish the available mass reservoir. 
\citet{kumar07} suggested that this morphology of filaments compare well with simulations relying on the competitive accretion scenario (\citealt{bate03}).  Actually, the results of the \citet{smith09b} simulations (see their Fig. 1), based on competitive accretion, shows a morphology very similar to the one found in DR 21, with gas filaments feeding the star formation in narrow filament-like structures. 


\begin{figure*}
\hspace{0cm}
\centering 
\includegraphics[angle=0,width=4.3cm]{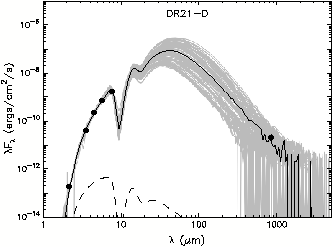}
\includegraphics[angle=0,width=4.3cm]{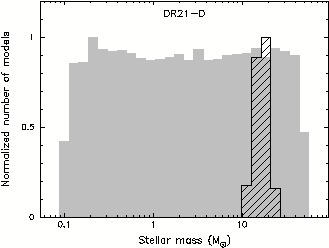}
\includegraphics[angle=0,width=4.3cm]{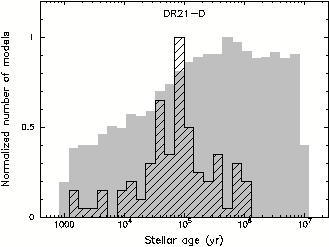}
\includegraphics[angle=0,width=4.3cm]{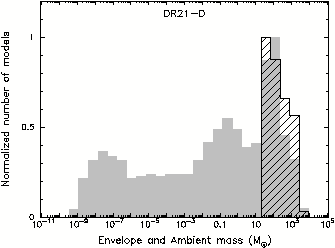}
   \caption{Results of the best 100 fits for the DR 21-D source obtaining by using the \citet{robitaille06} stellar models. From left to right: SED, stellar mass, stellar age and circumstellar envelope mass. In the SED panel, the filled circles show the input fluxes, the black line shows the best fit, and the gray lines show subsequent 99 best fits. The dashed line shows the stellar photosphere corresponding to the central source of the best fitting model, as it would be seen in the absence of circumstellar dust (but including interstellar extinction). In the 3 rightmost panels, the gray histogram shows the distribution of models in the model grid, and the hashed histogram shows the distribution of the 100 best models.}
\label{fig13bis}
\end{figure*}

Since the filaments are gravitationally unstable (\citealt{hennemann12}), the inflowing material can fragment (\citealt{csengeri11}) and form stars. As a result, a stellar filament is formed. Since the filaments point towards the accretion center, these filaments can exhibit a radial structure that resembles the spokes of a wheel. \citet{teixeira06} detected this "Spokes$-$like" structure in the stellar cluster NGC 2264, with the more massive stars located towards the center. The morphology of the stellar population in radial filaments is thought to represent primordial structures in the formation of stellar clusters (\citealt{bate03,kurosawa04,teixeira06}). We have detected in DR 21 a similar structure. In Section \ref{results} and Section \ref{angular} we showed that the X/noSDSS/IR samples exhibit an stellar filament from the DR 21 core towards the NE.
 
Once the gas fall towards the cloud, it can feed further stellar formation. The overall gravitational well will funnel a significant fraction of gas and dust to the DR 21 core (explaining the observed overall centrally peaked structure of the extinction), and favoring the formation of a dense low-mass stellar cluster and massive stars at the center, as observed in the DR 21 core. However, this do not rule out that massive stars can also born in outer parts of the cluster. The denser sub-clusters of low-mass stars can also benefit from their own small-scale gravitational potential wells, winning the competition from the surrounding mass reservoir and allowing the formation of massive stars. We propose that this mechanism could explain the massive star formation in the the DR 21(OH) and FIR 1/2/3 regions (which show secondary peaks in the extinction, see right panel of Fig. \ref{fig6}). 

In agreement with this scenario, the millimeter interferometric observations from \citet{bontemps10} showed that the dense condensations of gas that are expected to be the earliest stages of massive star evolution in the DR 21 cloud (CygX-N48 and CygX-N53, in the DR 21(OH) and FIR 1/2/3 regions, respectively) appears fragmented, forming sub-clusters. \citet{csengeri11} proposed that significant amount of competitive accretion may be present in these condensations. In this direction, \citet{rivilla13a} pointed out that the low-mass stellar population found in the massive star-forming regions in Orion strongly indicates that the natal condensations suffers high levels of fragmentation, forming sub-clusters of low-mass stars rather than single massive objects. 

A consequence of the proposed scenario for the DR 21 cluster formation is that one also would expect that the younger stars of the cluster are located along the overall potential gravitational well, where the material from the natal cloud is funneled via the filaments. We discuss here some aspects about the estimated ages of the DR 21 stellar population. 

\citet{prisinzano08} and \citet{ybarra13}, combining Chandra and Spitzer data, found a relation between the extinction of the stars and their evolutionary phase: the more extincted the star appears, the younger the star is. Given that the extinction is significantly higher along the N-S cloud, and in particular in the DR 21 core, this would imply an analogous distribution of stellar ages. \citet{kumar07} already pointed out that the Spitzer sources with higher IRAC spectral indexes (indicative of youth) are distributed along the dense cloud and also coincident with signposts of massive star formation. \citet{beerer10} also find that the younger Class I objects follow the cloud, while the Class II object are more distributed. Our classification of the X/noSDSS/UKIDSS/Spitzer sample also shows that the youngest sources are located within the molecular cloud (Section \ref{spitzer}). 

Using the relation between the extinction and the age found by \citet{ybarra13} in the Rossete Nebula, the visual extinction suffered by the stars more deeply embedded in the cloud ($A_{\rm V}>$15 mag, left panel of Fig. \ref{fig6}) implies that their expected age is $\sim$10$^{5}$ yr, in agreement with the estimated age for Class I objects (\citealt{andre94,evans09}). 

To check this we have determined quantitatively the age of one of the embedded stars in the cloud, DR 21-D\footnote{We note that this source is not included in the general analysis of this paper because Chandra did not detect emission, very likely due to the very high absorption produced by the dense circumstellar material.}. This source is located in the core of DR 21, believe to ionize the UC HII region D (\citealt{cyganowski03}). We fitted the spectral energy distribution of the source with the \citet{robitaille06} stellar models, using the $K$-band flux from UKIDSS, the Spitzer fluxes, and the SCUBA 850 $\mu$m flux. In Fig. \ref{fig13bis} we show the results of the best 100 models. We obtain that this object is a $\sim$10$-$11 M$_{\odot}$ star with a luminosity of $\sim$10$^{4}$ L$_{\odot}$, heavily obscured by $A_{\rm V}>$100, with a very massive circumstellar envelope and with an estimated stellar age of $\leq$10$^{5}$ yr. This is fully consistent with our previous estimate of the age of the stars deeply embedded within the N-S cloud. On the other hand, Fig. \ref{fig9} shows that more evolved (but still young) stars (Class II), with an estimated age of 10$^{6}$ yr, exhibit a more spatially distributed configuration. 
The NE stellar filament has a population of Class II and Class III objects, indicating that they a more evolved that the sources detected in the cloud (Fig. \ref{fig9}). This would imply that the stellar filament has already consumed a significant fraction of the initial gas, in agreement with the lower extinction we have found (Section \ref{UKIDSS} and Fig. \ref{fig7}). This also explain why the filament is not clearly detected at 8 $\mu$m, far-IR (\citealt{hennemann12}), 850 $\mu$m or by molecular observations. 
We note that the absence of the Class II/III objects along the N-S cloud could be caused by the higher extinction in this region. However, the non detection of Class I sources in the outer parts of the field clearly indicate that the more youngest sources are concentrated in the cloud. This age segregation would be consistent with the competitive accretion formation of the cluster, although needs to be confirmed by higher sensitivity observations at mid-IR, far-IR and submillimeter wavelengths of the individual stars. 

In summary, all the findings of this work and previous observations agree with the predictions of the scenario where massive star formation is directly linked to the formation and early evolution of the low-mass stellar clusters, and where competitive accretion plays a crucial role.

\begin{figure*}
\centering 
\hspace{-0.65cm}   
\includegraphics[angle=0,width=17cm]{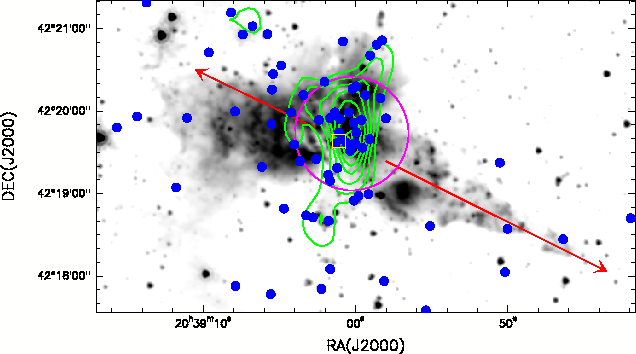}
   \caption{Zoom-in view of the DR 21 core and the large high-energetic outflow. The sources from the X/noSDSS sample are denoted with blue dots. The open yellow square indicates the position of the explosion center proposed by \citet{zapata13}, and the red arrows follow the H$_{2}$ outflow lobes observed by \citealt{davis07} (see their Fig. 2). The magenta circle indicate the DR 21 core cluster diameter (0.6 pc) calculated in Section \ref{stellar-density-map}. The green contours are the stellar density obtained with the 0.22 pc $\times$ 0.22 pc grid, as in Fig \ref{fig10}. The background gray scale image is the Spitzer 4.6 $\mu$m image, where the outflow is also detected.}
\label{fig14}
\end{figure*}

\section{A stellar collision as the origin of the large-scale outflow?}
\label{origin-outflow}

The Chandra detection of a dense stellar cluster of young stars in the core of DR 21, at the expected position of the explosive event proposed by \citet{zapata13} (see Fig.  \ref{fig14}), leads us to consider the possibility of a coalescence of stars as the origin of the highly-energetic outflow. 

\citet{zapata11} proposed that the large-scale outflow in Orion BN/KL region was produced by an explosive event. Indeed, this is a plausible scenario since \citet{rivilla13a} found that the stellar cluster density is $>$10$^{6}$ stars pc${-3}$, making the collision between two stars a likely event. Although the stellar density in the DR 21 region is about one order of magnitude lower than that found in the Orion BN/KL region, one may wonder whether a collision can also have occurred in the DR 21 core. A collision event in a dense cluster is favored by the presence of circumstellar disks via disk-captures, which enhance the probability of an encounter (\citealt{zinnecker07,davies06,bonnell05}). This opens the possibility of having stellar mergers that may generate a powerful outflow. 

\begin{figure}
   \centering 
\includegraphics[angle=0,width=8cm]{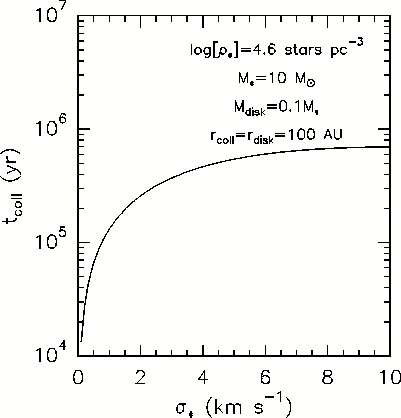}
   \caption{Time expected to produce a collision in the DR 21 dense stellar cluster, as a function of the stellar velocity dispersion $\sigma_{*}$. We used the stellar density found in the DR 21 inner region (log $\rho_{*}\sim$4.6 stars pc$^{-3}$), and assume a collision between two stars with $M_{*}$=10 M$_{\odot}$ stars with disk with masses 0.1 $M_{*}$ and radius of 100 AU.}
\label{fig15}
\end{figure}

\citet{rivilla13a} studied a collision involving disks and obtained an expression for the expected time of collision between two stars, $t_{\rm coll}$ (their Eq. A.4.). In Fig. \ref{fig15} we show $t_{\rm coll}$ versus the stellar velocity ($\sigma_{*}$). We used similar stellar parameters to those derived for DR 21 D (Section \ref{star-formation}), i.e., masses of M$_{*}$=10 M$_{\odot}$ with disks of 0.1 M$_{*}$ and radius of 100 AU. We considered the stellar density estimated in the region in Section \ref{stellar-densities}.
As indicated by \citet{bonnell06}, one might expect that the stars forming part of a 
$small-N$
 young cluster like that in the DR 21 core have a low-velocity dispersion. These authors proposed to use values as low as 0.4 km s$^{-1}$. \citet{gomez08} claimed that typical random motions of recently formed stars have a velocity dispersion of 1$-$2 km s$^{-1}$. As an example, the ONC has a measured stellar velocity dispersion of $\sim$2.3 km s$^{-1}$ (\citealt{vanaltena88}). With these values, Fig. \ref{fig15} shows that a single collision in the system can occur in $\sim$1$-$2 $\times$ 10$^{5}$ yr. Given that the expected age for the stellar cluster embedded in the DR 21 core is around this value, we propose that a collision may have occurred in the center of this dense cluster, which could have led to the explosive event described in \citet{zapata13} and possible origin of the large-scale DR 21 outflow.

The energy lost produced during the collision between the disks of two stars can lead to the decay of the system, producing at the end a direct collision between the two stars. In that case, the energy released would be  $E=GM_{*}M_{*}/r$ (\citealt{bally05}), where $G$ is the gravitational constant, $M_{*}$ is the stellar mass, and $r$ is the radius of the collision. At that moment the radius of the collision corresponds to twice the value of the stellar radius. Considering then two 10 M$_{\odot}$ stars with and $r$ = 2 R$_{*}\sim$ 80 R$_{\odot}$\footnote{We considered the radius of a 10 M$_{\odot}$ star with an age of 10$^{5}$ yr, extrapolating from the PMS models of \citet{siess00}.}, the energy produced by the collision would be $\sim$5 $\times$ 10$^{48}$ ergs, which is of the same order as that observed in the DR 21 core (\citealt{zapata13}). Therefore, such a violent event produced by the coalescence of two massive members of the dense DR 21 stellar cluster could occur, and this would explain the highly-energetic outflow observed in the region. 

The stellar collisions producing large-scale outflows that may have occurred in DR 21 and Orion BN/KL regions suggest that these events might be more frequent than previously thought during the earliest phases of formation of dense stellar cluster harboring massive stars.

\section{Summary and conclusions}
\label{summary}

We presented for the first time the results of X-ray Chandra observations of the massive star-forming region DR 21. They are summarized as follows:

 - The X-rays observations have revealed a new highly embedded population of PMS low-mass stars previously missed in observations at other wavelengths.
 
 - The spatial distribution of the young low-mass PMS stars emitting X-rays shows a central concentration around the DR 21 core. The densest sub-clusters of low-mass stars coincide with the regions of massive star formation: the DR 21 core, DR 21(OH) and the FIR 1/2/3 region. 

- The X/noSDSS/IR sample of stars shows a stellar filament from the DR 21 core towards the NE, resembling a "Spokes-like" structure.

- We obtained the structure of the extinction, which decreases when the distance to the axis of the N-S cloud increases, in agreement with dust observations. Moreover, we found that the extinction also globally decreases with respect to the distance to the cluster center located in the DR 21 core.

- We classified the X/noSDSS/UKIDSS/Spitzer PMS stars in different evolutionary phases. Consistently with previous works, we find evidences for an age segregation, with the younger population (Class I) embedded in the N-S cloud, and more evolved sources (Class II/III) more distributed along the NE stellar filament and the rest of the field. 

- We found that the low-mass stellar population of the full cluster appears globally mass-segregated, with a trend of more massive stars at the center.

- The high stellar density found in the stellar cluster embedded in the DR 21 core may have induced the coalescence of two massive stars. We propose that this event could be the origin of the large-scale and highly-energetic DR 21 outflow detected in the region.

Our findings are consistent with a picture where the evolution of the stellar cluster and massive star formation are regulated by competitive accretion. The gravitational potential well of the full cluster accretes large amounts of gas through dusty and molecular filaments. These infalling filaments can also fragment and form stars, producing stellar filaments like the one observed in DR 21. Since the stellar density (and hence the gravitational potential) is centrally peaked towards the DR 21 core, a significant fraction of the material is funneled towards the center, where the densest low-mass stellar cluster is detected. This would explain the observed structure of the extinction, that increases when the distance to the DR 21 core center decreases. However, this does not mean that massive stars can not be formed outside the DR 21 core, as observed. In addition to the full potential well, the dense sub-clusters of low-mass stars can act at smaller scales. Their own local gravitational potential wells contribute to $win$ the competition for the surrounding gas reservoir, allowing the formation of massive stars in the DR 21(OH) and FIR 1/2/3 regions.

\section*{Acknowledgments}
We thank the anonymous referee, whose detailed and thoughtful comments and suggestions helped to improve significantly the original version of this paper.
This work  has been partially funded by MICINN grants AYA2010$-$21697$-$C05$-$01 and FIS2012$-$39162$-$C06$-$01, and Astro$-$
Madrid (CAM S2009/ESP$-$1496), and CSIC grant JAE$-$Predoc2008. I.J$-$S. acknowledges the financial support received from the People Programme (Marie Curie Actions) of the European Union's Seventh Framework Programme (FP7/2007$-$2013) under REA grant agreement number PIIF$-$GA$-$2011$-$301538.

\bibliographystyle{mn2e}
\bibliography{bib_DR21}

\bsp

\label{lastpage}

\end{document}